\documentclass[aps,preprint,prd,showpacs,nofootinbib]{revtex4}

\usepackage{amsmath,amssymb}
\usepackage{graphicx,subfigure}
\usepackage{color,multirow}
\usepackage[colorlinks,linkcolor=magenta,anchorcolor=cyan,citecolor=blue,plainpages=false]{hyperref}

\hypersetup{colorlinks=true,
    breaklinks=true,
    pdfstartview=Fit,
    linkcolor=blue,
    citecolor=blue,
    urlcolor=blue}

\def\be{\begin{equation}}
    \def\ee{\end{equation}}
\def\ba{\begin{eqnarray}}
    \def\ea{\end{eqnarray}}

\begin{document}
\title{Dark energy after pre-recombination early dark energy in light of DESI DR2 and the latest ACT and SPT data}

    \author{Hao Wang$^{1,2} $\footnote{\href{wanghao187@mails.ucas.ac.cn}{wanghao187@mails.ucas.ac.cn}}}
    \author{Yun-Song Piao$^{1,2,3,4} $ \footnote{\href{yspiao@ucas.ac.cn}{yspiao@ucas.ac.cn}}}

    \affiliation{$^1$ School of Fundamental Physics and Mathematical
        Sciences, Hangzhou Institute for Advanced Study, UCAS, Hangzhou
        310024, China}

    \affiliation{$^2$ School of Physical Sciences, University of
        Chinese Academy of Sciences, Beijing 100049, China}

    \affiliation{$^3$ International Center for Theoretical Physics
        Asia-Pacific, Beijing/Hangzhou, China}

    \affiliation{$^4$ Institute of Theoretical Physics, Chinese
        Academy of Sciences, P.O. Box 2735, Beijing 100190, China}

    \begin{abstract}

It has been noted that with the pre-recombination early dark
energy (EDE) resolution of Hubble tension, the preference of
recent datasets for the evolving dark energy (DE) can be
suppressed significantly. In this work, we clarify and reconfirm
this result with DESI DR2 and the latest ACT DR6 and SPT-3G D1,
the tightest small-scale CMB constraints up to date. In the
$w_0w_a$CDM model with EDE, a quintessence-like component
($w_0+w_a\geq-1$) can be 1$\sigma$ consistent with
Planck+ACT+SPT+DESI+Pantheon+SH0ES datasets, and
$\Delta\chi^2\lesssim -14$ compared with $w_0w_a$CDM model without
EDE. This reveals the possibility that when the potential
resolutions of Hubble tension are considered, current accelerated
expansion can attribute to a canonical evolving scalar field or
cosmological constant, and again highlights the importance of
re-examining the nature of DE within the broader context of
cosmological tensions.

    \end{abstract}

    \maketitle

\section{Introduction}

Despite its success in explaining broad cosmological observations,
the standard $\Lambda$CDM model is challenged by several issues,
most notably the Hubble tension
\cite{Verde:2019ivm,Riess:2019qba}. This tension arises from the
stark contrast between the Hubble constant measurements of $H_0
\sim 73$ km/s/Mpc from the SH0ES collaboration, which uses
Cepheid-calibrated Type Ia supernovae \cite{Breuval:2024lsv}, and
$H_0 \sim 67$ km/s/Mpc inferred by the Planck collaboration based
on the standard $\Lambda$CDM model using their cosmic microwave
background (CMB) data \cite{Planck:2018vyg}.

The Hubble tension has motivated many solution
attempts\cite{Knox:2019rjx,Perivolaropoulos:2021jda,DiValentino:2021izs,Vagnozzi:2023nrq},
among which Early Dark Energy (EDE)
\cite{Karwal:2016vyq,Poulin:2018cxd,Smith:2019ihp} is a leading
pre-recombination scenario. By becoming non-negligible just only
before recombination, EDE suppressed the sound horizon and raises
$H_0$, with an anti-de Sitter phase around recombination further
enhancing its effect (AdS-EDE) \cite{Ye:2020btb,Ye:2020oix}.
Though this uplift of $H_0$ often worsens the $S_8$ tension
\cite{Hill:2020osr,Ivanov:2020ril,DAmico:2020ods,Krishnan:2020obg,Nunes:2021ipq},
independent mechanisms may alleviate this,
e.g.\cite{Poulin:2022sgp,Allali:2021azp,Ye:2021iwa,Alexander:2022own,FrancoAbellan:2021sxk,Clark:2021hlo,Simon:2022ftd,Reeves:2022aoi,Wang:2022bmk,Chacko:2016kgg,Buen-Abad:2022kgf,Teng:2021cvy,Dainotti:2021pqg,Dainotti:2022bzg,Montani:2024pou}.

In the past decades, identifying the nature of DE has been still
an important challenge, see
e.g.\cite{Carroll:2000fy,Frieman:2008sn,Nojiri:2017ncd,Huterer:2017buf}
for reviews. According to the $w_0w_a$CDM
parametrization\cite{Chevallier:2000qy,Linder:2002et}, recent DESI
data\cite{DESI:2024lzq,DESI:2024mwx,DESI:2024uvr,DESI:2025zgx}
combined with CMB and supernovae data seems to prefer the evolving
dark energy against the cosmological constant (CC) at $\gtrsim
3\sigma$ significance level, challenging the $\Lambda$CDM model.
These results are under active investigation,
e.g.\cite{Luongo:2024fww,Cortes:2024lgw,Carloni:2024zpl,Colgain:2024xqj,Giare:2024smz,Wang:2024dka,Yang:2024kdo,Park:2024jns,Shlivko:2024llw,Dinda:2024kjf,Seto:2024cgo,Bhattacharya:2024hep,Roy:2024kni,Wang:2024hwd,Notari:2024rti,Heckman:2024apk,Gialamas:2024lyw,Orchard:2024bve,Colgain:2024ksa,Wang:2024sgo,Li:2024qso,Ye:2024ywg,Giare:2024gpk,Dinda:2024ktd,Jiang:2024viw,Alfano:2024jqn,Jiang:2024xnu,Sharma:2024mtq,Ghosh:2024kyd,Reboucas:2024smm,Pang:2024qyh,Wolf:2024eph,RoyChoudhury:2024wri,Arjona:2024dsr,Wolf:2024stt,Giare:2024ocw,Wang:2024tjd,Alestas:2024eic,Carloni:2024rrk,Bhattacharya:2024kxp,Specogna:2024euz,Li:2024qus,Ye:2024zpk,Pang:2024wul,Akthar:2024tua,Colgain:2024mtg,daCosta:2024grm,Chan-GyungPark:2025cri,Sabogal:2025mkp,Du:2025iow,Ferrari:2025egk,Jiang:2025ylr,Peng:2025nez,Jiang:2025hco,Feng:2025mlo,Hossain:2025grx,Chakraborty:2025syu,Borghetto:2025jrk,Pan:2025psn,Pang:2025lvh,Wang:2025ljj,Kessler:2025kju,Yang:2025mws,Wolf:2025jed,RoyChoudhury:2025dhe,Specogna:2025guo,Ye:2025ark,Cheng:2025lod,Ling:2025lmw,Wolf:2023uno,Wang:2025dtk,Li:2025eqh,Cline:2025sbt,Gialamas:2025pwv,Li:2025dwz,Lee:2025pzo,Ishak:2025cay,Wang:2025vtw,Wang:2025znm,Zhou:2025nkb,RoyChoudhury:2025iis,Pedrotti:2025ccw}.
However, it has been found that when the pre-recombination
resolution of Hubble tension is considered, this preference for
the evolving DE can be suppressed to $\lesssim2\sigma$
\cite{Wang:2024dka}, see also \cite{Pang:2024wul,Pang:2025lvh}.

The latest ACT DR6 and SPT-3G D1
data\cite{ACT:2025fju,ACT:2025tim,SPT-3G:2025bzu,Balkenhol:2024sbv}
are the most precise measurements of small-scale CMB polarization
to date, which has yielded the tightest CMB constraints on the
oscillations in primordial perturbation spectrum,
e.g.\cite{Peng:2025vda}. It is therefore timely and crucial to
investigate the implications of EDE models for the evolving DE in
light of latest ACT and SPT data. In this work, we consider two
representative EDE models, axion-like EDE
\cite{Karwal:2016vyq,Poulin:2018cxd,Smith:2019ihp} and AdS-EDE
\cite{Ye:2020btb,Ye:2020oix}, and find that the pre-recombination
resolutions of Hubble tension would still suppress the $\gtrsim
3\sigma$ evidence of DESI for the evolving DE with the latest ACT
and SPT data, so that a quintessence-like DE component
($w_0+w_a\geq-1$) can be 1$\sigma$ consistent.

\section{Why EDE prefers $w_=-1$?}

It is necessary to clarify why the $\gtrsim 3\sigma$ evidence of
DESI for the evolving DE is suppressed to $\lesssim2\sigma$ by the
pre-recombination resolution of Hubble tension.

In $\Lambda$CDM, the angular scale \be
\theta_s^*=\frac{r_s^*}{D_A^*}\label{thetas} \ee where $r_s^*$ is
the sound horizon at recombination and
$D_A^*=\int_{0}^{z^*}H^{-1}(z)dz=H_0^{-1}\int_{0}^{z^*}
E^{-1}(z)dz$ is the angular diameter distance to the last
scattering surface, has been precisely measured by the CMB. In
light of the poor effect of modifying the low-redshift evolution
$E(z)$, to make $H_0$ higher but not spoil the fit to CMB, a
smaller $r_s^*$, so EDE, is required.

It is known that $D(z)/r_d$ is constrained by DESI. As a result
the degeneracy between $\Omega_m$ and $H_0r_d$ is captured, and
\be \int_{0}^{z^*}
E^{-1}(z)dz\equiv\int_{0}^{z^*}\frac{dz'}{\sqrt{\Omega_m(1+z)^3+\Omega_{\rm
DE}(z)}}=\frac{H_0r_d}{\theta_s^*} \ee is required to be a
constant to reproduce the $\Lambda$CDM-like result. Thus if the DE
is evoling, its $w(z)$ must be larger than $-1$ at one period
while smaller than $-1$ in other period such that the mean effect
recovers $w=-1$. This naturally leads to a non-negligible
probability that the phantom crossing occurs
\cite{Ye:2024ywg,DESI:2024aqx,Ozulker:2025ehg,Shlivko:2024llw,Gialamas:2025pwv,Keeley:2025rlg,Cai:2025mas},
However, the pre-recombination scenarios have potential effect on
the low-redshift evolution. According to (\ref{thetas}), although
changing $E(z)$ has little benefit on lifting $H_0$, the
suppression of $r_s^*$ caused by EDE could change not only $H_0$
but $E(z)$.

The Hubble parameter $H(z)$ is also relevant to the nature of DE,
and for the parameterization $w(z)=w_0+w_az/(1+z)$, we have
\begin{widetext}
    \begin{equation}
        H(z)/H_0=\left[\Omega_m(1+z)^3+(1-\Omega_m)(1+z)^{3(1+w_0+w_a)}e^{-3w_az/(1+z)}\right]^{1/2}.
        \label{HzH0}
    \end{equation}
\end{widetext}
The comoving distances $D_M(z)\equiv\int_{0}^{z}{cdz'/H(z')}$ and
$D_H(z)\equiv{c/ H(z)}$ are independently constrained by DESI,
thus a high $H_0$ will help to suppress the requirements of data
for $w_0>-1$ and $w_a<0$, and make both closer to $w_0=-1$ and
$w_a=0$.

\section{Testing DE after pre-recombination EDEs}

\subsection{Data and Methods}

We combine the \texttt{ACT-lite} likelihood for ACT
DR6\cite{ACT:2025fju,ACT:2025tim} + \texttt{SPT-lite} likelihood
for SPT3G D1\cite{SPT-3G:2025bzu,Balkenhol:2024sbv} +
\texttt{Plik-lite} likelihood cut at $l>1000$ in TT, and  $l>600$
in TE and EE and the Planck \texttt{Commander} and \texttt{SimALL}
likelihood for low-$l$ TT and EE spectra\cite{Planck:2019nip}, as
well as the CMB lensing data of Planck PR4\cite{Carron:2022eyg},
ACT DR6\cite{ACT:2023dou,ACT:2023kun,ACT:2023ubw} and
SPT-3G\cite{SPT-3G:2024atg,SPT-3G:2025zuh}. We denoted the
Planck+ACT+SPT datasets as $\mathbf{SPA}$. We also include the
$\mathbf{DESI}$ DR2 BAO data\cite{DESI:2025zgx}, the uncalibrated
Type Ia SN from the $\mathbf{Pantheon}$
dataset\cite{Scolnic:2021amr} and the SH0ES Cepheid calibrated
$\mathbf{Pantheon+SH0ES}$\cite{Riess:2021jrx}, respectively.

In the $w_0w_a$CDM+EDE scenario, e.g.\cite{Wang:2022jpo}, the
equation of state of DE considered is the CPL parametrization
\cite{Chevallier:2000qy,Linder:2002et}
\begin{equation}
    w(z)=w_0+w_a\frac{z}{1+z}.
\end{equation}
One of EDEs considered is axion-like EDE with an axion-like
potential \cite{Poulin:2018cxd,Poulin:2018dzj}:
$V(\phi)=m^2f^2[1-\cos(\phi/f)]^3$,
where $m$ and $f$ are the effective mass and the couple constant
of EDE, respectively. The other is AdS-EDE
\cite{Ye:2020btb,Ye:2020oix,Ye:2022efx,Jiang:2021bab}, in which
\begin{equation}\label{Vede}
    V(\phi)=\left\{\begin{array}{ll}
V_{0}\left(\dfrac{\phi}{M_{{p}}}\right)^{4}-V_\mathrm{ads},\quad \dfrac{\phi}{M_{{p}}}<\left(\dfrac{V_\mathrm{ads}}{V_{0}}\right)^{1 / 4} \\
    0, \quad\quad\quad\quad\quad\quad\quad\quad \dfrac{\phi}{M_{{p}}}>\left(\dfrac{V_\mathrm{ads}}{V_{0}}\right)^{1 / 4}
    \end{array}\right.
\end{equation}
where $V_\mathrm{ads}$ is the AdS depth, which is fixed by
$\alpha_{ads}\equiv
{V_\mathrm{ads}/\rho_m(z_c)+\rho_r(z_c)}=3.79\times10^{-4}$ as an
effective shortcut to avoid bad convergence of the
chain\cite{Ye:2020btb}.

We perform the Markov chain Monte Carlo (MCMC) analysis using
$\mathbf{Cobaya}$\cite{Torrado:2020dgo} and the cosmological
Boltzmann code $\mathbf{CLASS}$\cite{Blas:2011rf}. In our MCMC,
the parameter sets are $\{\omega_b, \omega_{cdm}, H_0,
\ln10^{10}A_s, n_s, \tau_{reio}, w_0, w_a, \log_{10}z_c,
f_\mathrm{EDE}, \theta_i\}$ for axion-like EDE, where
$\theta_i\equiv \phi_i/f$ and $f_\mathrm{EDE}$ is the energy
fraction of EDE at $z_c$, and $\{\omega_b, \omega_{cdm}, H_0,
\ln10^{10}A_s, n_s, \tau_{reio}, w_0, w_a, \log_{10}z_c,
f_\mathrm{EDE}\}$ for AdS-EDE. We adopt the Gelman-Rubin
convergence criterion with a threshold
$R-1<0.01$\cite{Gelman:1992zz}.

\subsection{Results}

\begin{figure*}
    \includegraphics[width=1\columnwidth]{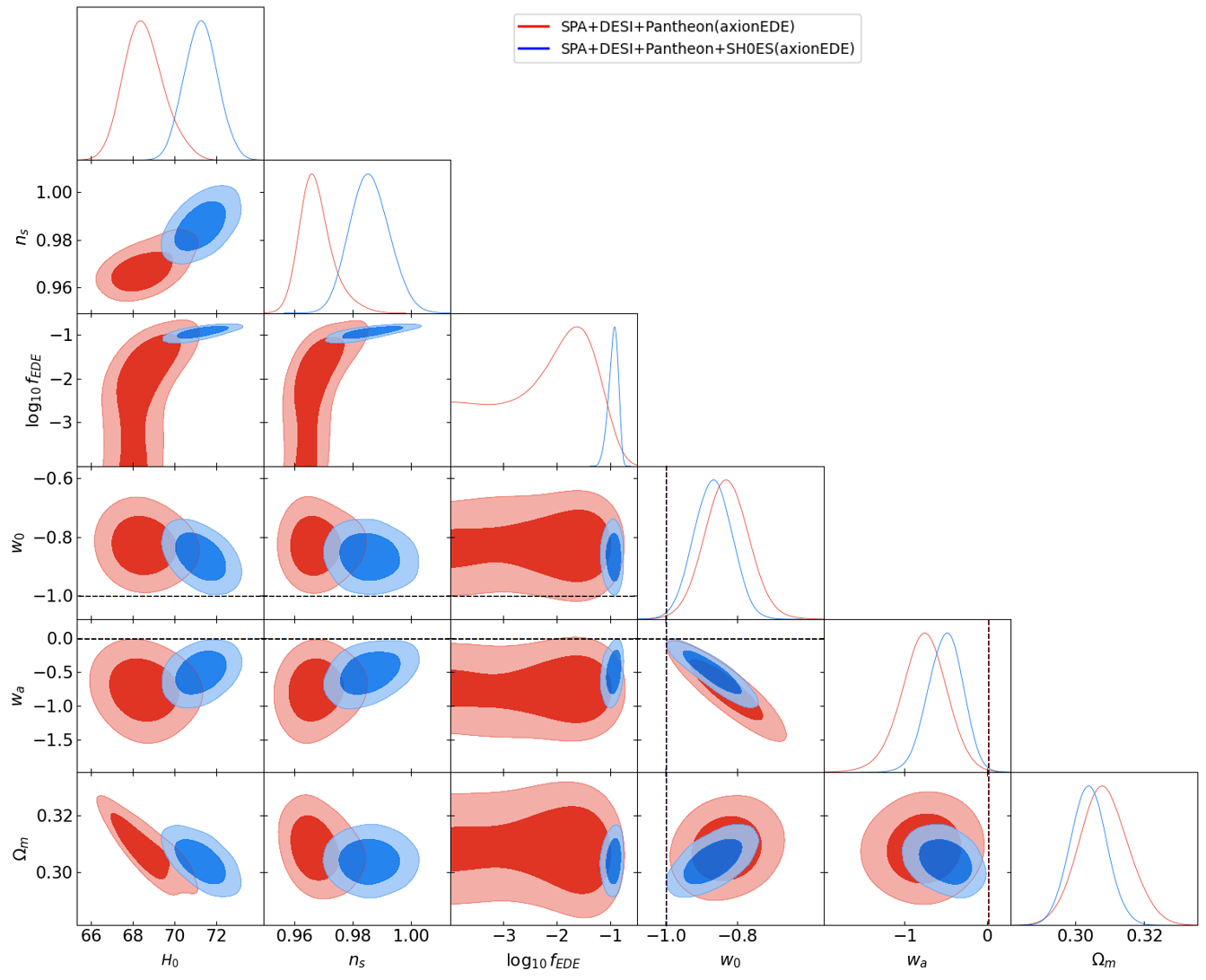}
    \caption{\label{axi}2D contours of the
            primary parameters at 68\% and 95\% CL for the
            $w_0w_a$CDM+axionlike EDE model.}
\end{figure*}

\begin{figure*}
    \includegraphics[width=1\columnwidth]{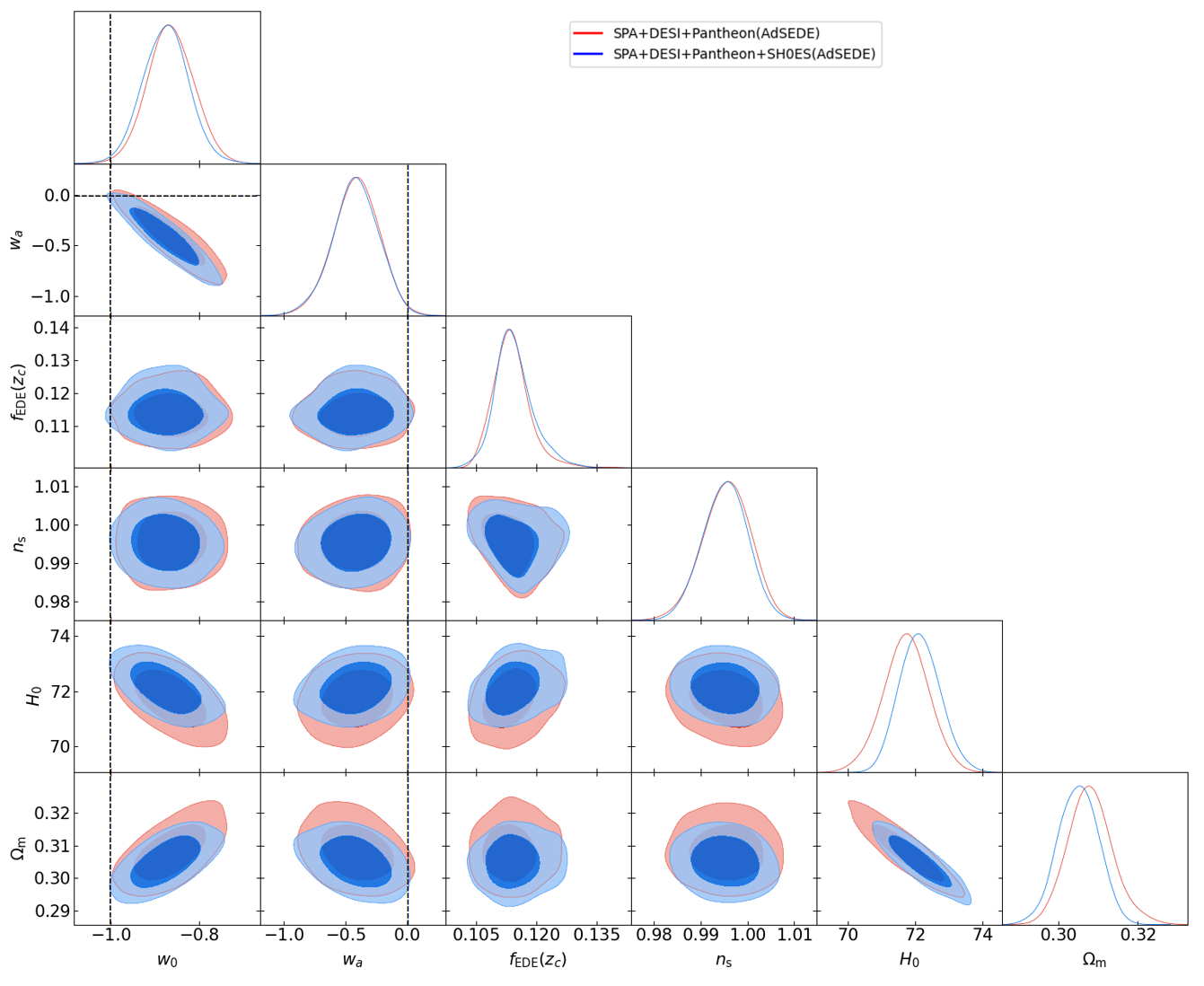}
    \caption{\label{ads}2D contours of the
            primary parameters at 68\% and 95\% CL for the
            $w_0w_a$CDM+AdSEDE model.}
\end{figure*}

The mean (bestfit) values and 1$\sigma$ errors of cosmological
parameters are presented in Tables \ref{tab}. The 2D posteriors
are displayed in Fig.\ref{axi} and Fig.\ref{ads} for axion-like
EDE and AdS-EDE, respectively.

The result for axion-EDE with SH0ES is $w_0=-0.868\pm0.053$ and
$w_a=-0.507\pm0.208$, and CC is $\simeq2\sigma$ consistent. That
for AdS-EDE is $w_0=-0.875\pm0.052$ and $w_a=-0.416\pm0.188$,
which indicates $\lesssim2\sigma$ significance for evolving DE.
This is similar to the earlier result in Ref.\cite{Wang:2024dka}
using Planck 2018 CMB data and DESI DR1. However, even if not
including SH0ES there is still a only $\simeq2\sigma$ significance
for evolving DE in both EDE models, as displayed in Fig.\ref{axi}
and Fig.\ref{ads}.
We also notice that both
axion-like EDE and AdS-EDE show a lower $H_0$ with ACT+SPT than
the earlier result in Ref.\cite{Wang:2024dka},
which attributes to the tighter constraint on $f_\mathrm{EDE}$ by
Planck+ACT+SPT, as pointed in Ref.\cite{Peng:2025tqt}.

We list the $\chi^2$ values in Table \ref{chi2}. It is significant
to see that for SPA+DESI+Pantheon, we have
$\Delta\chi^2\simeq-8.2$ and $\Delta\chi^2\simeq-2.1$ for
axion-like EDE and AdS-EDE with $w_0w_a$CDM compared with
$w_0w_a$CDM model, respectively, while for
SPA+DESI+Pantheon+SH0ES, we have $\Delta\chi^2\simeq-17.5$ and
$\Delta\chi^2\simeq-14.1$. The analysis combined with ACT+SPT and
evolving DE strengthens the significance of EDE on resolving
Hubble tension. It is also interesting to note that DESI data
prefer AdS-EDE at $\Delta\chi^2_\mathrm{DESI}\simeq-2.3$ without
SH0ES and $\Delta\chi^2_\mathrm{DESI}\simeq-1.5$ with SH0ES, but
the AdS-EDE model slightly exacerbate on the fitting to ACT DR6
data.


   \begin{table}[htbp]
    \centering
    \begin{tabular}{c|c|c|c|c}
        \hline  \multirow[c]{2}{*}{Parameters}& \multicolumn{2}{c|}{axion-like EDE} & \multicolumn{2}{c}{AdS-EDE}\\
        \cline{2-5}
        &w/o SH0ES&w/t SH0ES& w/o SH0ES&w/t SH0ES\\
        \hline
        $100\omega_b$&2.238(2.265)$\pm$0.021&2.251(2.249)$\pm$0.018&2.306(2.308)$\pm$0.145&2.305(2.306)$\pm$0.013\\
        $\omega_{cdm}$&0.121(0.126)$\pm$0.002&0.131(0.132)$\pm$0.002&0.134(0.133)$\pm$0.001&0.135(0.133)$\pm$0.001\\
        $H_0$&68.53(69.97)$\pm$0.95&71.27(71.88)$\pm$0.79&71.72(71.83)$\pm$0.68&72.14(71.91)$\pm$0.60\\
        $\ln10^{10}A_s$&3.041(3.035)$\pm$0.014&3.065(3.064)$\pm$0.014&3.057(3.060)$\pm$0.014&3.056(3.057)$\pm$0.014\\
        $n_s$&0.967(0.977)$\pm$0.005&0.985(0.988)$\pm$0.006&0.995(0.993)$\pm$0.005&0.995(0.993)$\pm$0.004\\
        $\tau_{reio}$&0.053(0.047)$\pm$0.007&0.056(0.055)$\pm$0.007&0.045(0.046)$\pm$0.008&0.044(0.045)$\pm$0.008\\
        \hline
        $w_0$&-0.829(-0.813)$\pm$0.060&-0.868(-0.854)$\pm$0.053&-0.864(-0.907)$\pm$0.052&-0.875(-0.885)$\pm$0.052\\
        $w_a$&-0.764(-0.803)$\pm$0.264&-0.507(-0.575)$\pm$0.208&-0.411(-0.234)$\pm$0.187&-0.416(-0.324)$\pm$0.188\\
        \hline
        $f_\mathrm{EDE}$&$<0.102$&0.114(0.103)$\pm$0.024&0.113(0.110)$\pm$0.004&0.114(0.111)$\pm$0.004\\
        $\log_{10}z_c$&3.471(3.512)$\pm$0.150&3.542(3.576)$\pm$0.053&3.501(3.489)$\pm$0.036&3.497(3.479)$\pm$0.035\\
        $\theta_i$&2.721(2.843)$\pm$0.110&2.780(2.769)$\pm$0.092&-&-\\
        \hline
        $\Omega_m$&0.308(0.306)$\pm$0.007&0.303(0.300)$\pm$0.005&0.308(0.304)$\pm$0.005&0.305(0.304)$\pm$0.005\\
        $S_8$&0.834(0.836)$\pm$0.010&0.846(0.845)$\pm$0.009&0.858(0.849)$\pm$0.007&0.858(0.851)$\pm$0.007\\
        \hline
    \end{tabular}
\caption{\label{tab}Mean(bestfit) values and 1$\sigma$ regions of
the parameters for the axion-like EDE and AdS-EDE models. The
dataset used is the $\mathbf{SPA+DESI+Pantheon}$(with and without
SH0ES) dataset.}
\end{table}

    \begin{table}[htbp]
    \centering
    \begin{tabular}{c|c|c|c|c|c|c}
        $\chi^2$&\multicolumn{2}{c|}{$w_0w_a$CDM}&\multicolumn{2}{c|}{AdSEDE}&\multicolumn{2}{c}{axionEDE}\\
        \hline
        Planck2018 low-$l$ TT&22.37&22.41&21.03&21.13&21.20&21.07\\
        Planck2018 low-$l$ TT&397.37&397.43&395.23&395.33&395.60&395.50\\
        PlanckActCut&220.04&220.96&222.79&222.97&217.71&218.63\\
        ACTDR6&158.90&159.12&157.95&157.97&157.20&157.70\\
        CMB lensing&37.48&38.71&38.68&38.49&38.09&38.51\\
        SPT3G&171.09&171.50&171.63&171.77&170.33&170.61\\
        DESI&11.48&10.60&9.14&9.02&10.81&9.90\\
        Pantheon&1402.39&-&1402.62&-&1402.33&-\\
        Pantheon+SH0ES&-&1420.17&-&1403.14&-&1404.48\\
        \hline
        $\chi_{\mathrm{total}}^2$&2421.12&2433.90&2419.09&2419.74&2412.93&2416.40\\
    \end{tabular}
\caption{\label{chi2}$\chi^2$ of $w_0w_a$CDM and EDE models
fitting to $\mathbf{SPA+DESI+Pantheon}$ dataset, with and without
SH0ES.}
\end{table}

\section{Conclusion}\label{chp4}

In this paper we accessed the impact of DESI DR2 and lastest
ACT+SPT data, the current most precise small-scale measurements of
CMB, on the state equation of DE in potential Hubble-tension-free
cosmologies. We find that the pre-recombination resolution of
Hubble tension significantly suppress the $\gtrsim 3\sigma$
evidence of DESI for the evolving DE, though $w_0>-1$ and $w_a<0$
are still hinted.
This reconfirm our earlier result \cite{Wang:2024dka} but with
DESI DR2 and latest ACT+SPT data. According to our result, not
considering the Hubble tensions which the $\Lambda$CDM and/or
$w_0w_a$CDM model is suffering from might bias our insight into
the nature of DE.

\begin{figure*}
    \includegraphics[width=0.45\columnwidth]{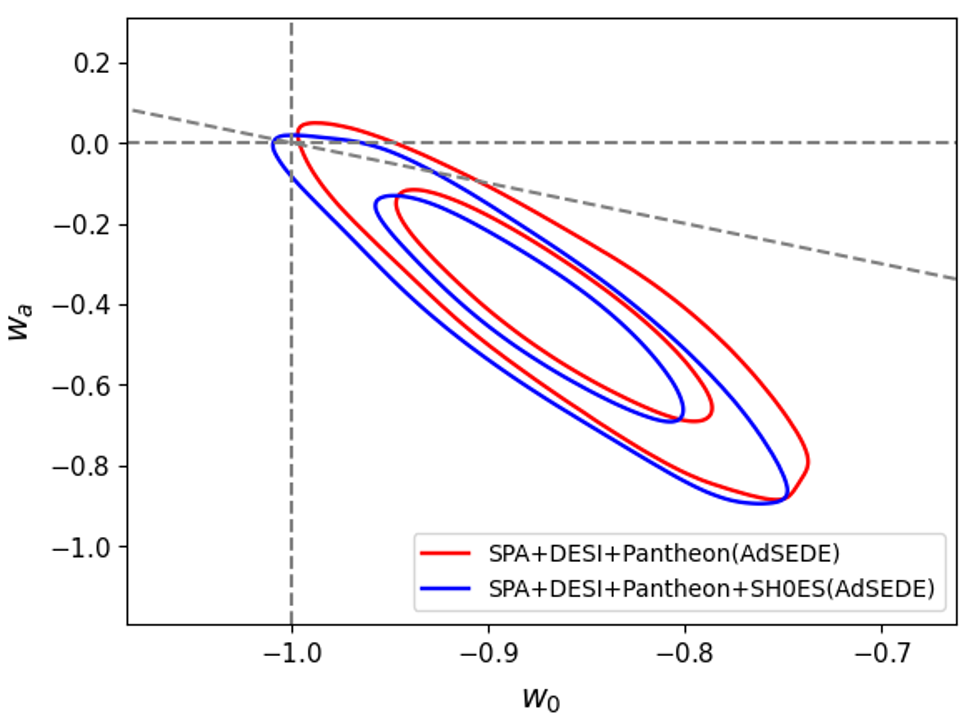}
    \includegraphics[width=0.45\columnwidth]{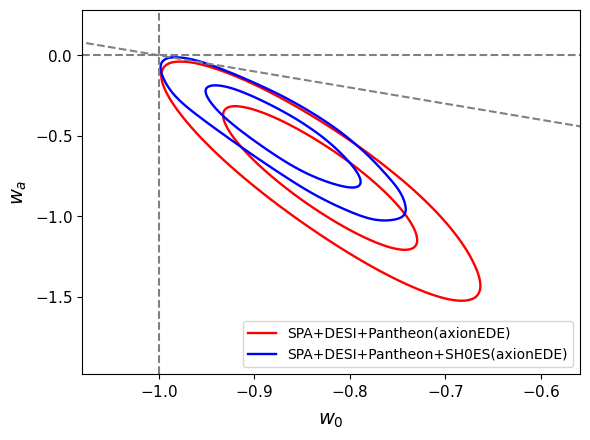}
\caption{\label{phan}2D contours of the $w_0-w_a$ parameters at
68\% and 95\% CL for the $w_0w_a$CDM+EDE models. The inclined
dashed grey lines depict the phantom-divided bound. It can be seen
that with AdS-EDE the quintessence model without the
phantom-divided crossing can be still $\sim 1\sigma$ consistent. }
\end{figure*}

The combination of CMB+DESI+SN seems to suggest not only an
evolving DE component but also one with the state parameter
crossing the phantom divide line ($w_0+w_a=-1$), which is not just
present within the CPL parametrization but is also seen in studies
reconstructing the DE density and the equation of state parameter
of DE in a model-independent way, which has been investigated in
Refs.\cite{Ye:2024ywg,DESI:2024aqx,Ozulker:2025ehg,Shlivko:2024llw,Gialamas:2025pwv,Keeley:2025rlg,Cai:2025mas}.
Combined with the pre-recombination resolution of Hubble tension,
we find a quintessence-like component ($w_0+w_a\geq-1$) can be
1$\sigma$ consistent with recent CMB+DESI+SN dataset, even
including the latest ACT+SPT CMB data, see Fig.\ref{phan}. This
reveals the possibility that with the pre-recombination EDE,
current accelerated expansion can attribute to a canonical
evolving scalar field or CC, and again highlights the importance
of re-examining our understanding on DE within the broader context
of cosmological tensions.

It is well known that inflation predicts nearly scale-invariant
scalar perturbation. Though based on the $\Lambda$CDM model the
Planck collaboration inferred the scalar spectral index is
$n_s\approx 0.965$ \cite{Planck:2018jri}, when the EDE resolution
of Hubble tension is considered, $n_s=1$ ($n_s-1\sim {\cal O}
(0.001)$) is favored
\cite{Ye:2020btb,Ye:2021nej,Jiang:2022uyg,Smith:2022hwi}, see also
\cite{Jiang:2022qlj,Peng:2023bik,DiValentino:2018zjj,Giare:2022rvg,Kallosh:2022ggf,Braglia:2020bym,Ye:2022efx,Jiang:2023bsz,Braglia:2022phb,DAmico:2021fhz,Takahashi:2021bti,Giare:2023wzl,Fu:2023tfo,Fu:2025ciy,Giare:2024akf,Peng:2025tqt}
for relevant studies on inflation models. Here, with DESI and
ACT+SPT data we find that $n_s=1$ persists, see Table.\ref{tab}
and see also \cite{Peng:2025tqt}. It will be interesting to
revisit inflation models, or pre-inflationary physics
e.g.\cite{Piao:2003zm,Piao:2005ag,Piao:2004hr,Piao:2003ty,Liu:2011ns},
in the light of our current constraints on $n_s$.

\section*{Acknowledgments}
This work is supported by NSFC, No.12075246, National Key Research
and Development Program of China, No.2021YFC2203004, and the
Fundamental Research Funds for the Central Universities.

\end{document}